**Orbital occupancy and charge doping in iron-based superconductors**

*Claudia Cantoni\*, Jonathan E. Mitchell, Andrew F. May, Michael A. McGuire, Juan-Carlos Idrobo, Tom Berlijn, Elbio Dagotto, Matthew F. Chisholm, Wu Zhou, Stephen J. Pennycook, Athena S. Sefat, and Brian C. Sales*

Dr. C. Cantoni, Mr. J. E. Mitchell, Dr. A. F. May, Dr. A. S. Sefat, Dr. M. A. McGuire, Dr. J.-C. Idrobo, Dr. Matthew F. Chisholm, Dr. Stephen J. Pennycook, and Dr. Brian C. Sales
Materials Science and Technology Division, Oak Ridge National Laboratory,
Oak Ridge TN, 37831, USA
E-mail: cantonic@ornl.gov
Dr. J.-C. Idrobo, Dr. W. Zhou
Department of Physics and Astronomy, Vanderbilt University, Nashville TN, 37235, USA
Dr. Berlijn
Center for Nanophase Materials Sciences, Oak Ridge National Laboratory, Oak Ridge, TN 37831, USA
Prof. E. Dagotto
Department of Physics and Astronomy, University of Tennessee, Knoxville, TN 37996, USA

**Iron-based superconductors (FBS) comprise several families of compounds having the same atomic building blocks for superconductivity, but large discrepancies among their physical properties. A longstanding goal in the field has been to decipher the key underlying factors controlling $T_C$ and the various doping mechanisms. In FBS materials this is complicated immensely by the different crystal and magnetic structures exhibited by the different families. In this paper, using aberration-corrected scanning transmission electron microscopy (STEM) coupled with electron energy loss spectroscopy (EELS), we observe a universal behavior in the hole concentration and magnetic moment across different families. All the parent materials have the same total number of electrons in the Fe 3$d$ bands; however, the local Fe magnetic moment varies due to different orbital occupancy. Although the common understanding has been that both long-range and local magnetic moments decrease with doping, we find that, near**

**the onset of superconductivity, the local magnetic moment increases and shows a dome-like maximum near optimal doping, where no ordered magnetic moment is present. In addition, we address a longstanding debate concerning how Co substitutions induces superconductivity in the 122 arsenide family, showing that the 3d band filling increases a function of doping. These new microscopic insights into the properties of FBS demonstrate the importance of spin fluctuations for the superconducting state, reveal changes in orbital occupancy among different families of FBS, and confirm charge doping as one of the mechanisms responsible for superconductivity in 122 arsenides. More generally, here we establish the validity of a method for comparing local magnetic moments that can be adopted for many other classes of Fe- and transition-metal-compounds, and used in the future to map the local magnetic moment with atomic spatial resolution. This will allow a deeper understanding of magnetic phenomena in technologically interesting materials.**

Extensive experimental and theoretical work has been carried out in recent years to gain a comprehensive understanding of the complex electronic structure of iron-based superconductors (FBS). Although a qualitative picture for pnictides has initially evolved based on common features, such as Fermi surface nesting between electron and hole pockets, and small magnetic moments in the paramagnetic state,[1-3] recent results for chalcogenides, i.e. $A_y$Fe$_{2-x}$Se$_2$ (AFS; $A$ = alkali metal or alkaline-earth metal), have challenged those conclusions. The AFS have the largest magnetic moments among FBS and lack hole pockets in the Fermi surface.[4,5] Given the large differences in the electronic structure among FBS families, identifying the electronic features relevant for superconductivity has been a formidable challenge, making FBS particularly puzzling materials.

One of the most controversial aspects of FBS is their magnetism. Initially, the ordered magnetic state in FBS was described as a spin-density-wave related to the Fermi surface nesting and originating from itinerant electrons.[6] Currently, it is widely accepted that magnetic correlations in FBS cannot be solely explained by either a pure weak-coupling approach based on Fermi surface nesting, or a pure strong-coupling approach based on local moments.[7, 8] AFS with composition $A_2Fe_4Se_5$ have the largest local magnetic moment (LMM), and for these materials, different experimental techniques as well as theoretical calculations agree on a magnitude of ~ 3.3 $\mu_B$. For the other FBS, in particular pnictides, the situation is confusing. First-principles calculations, based on density functional theory (DFT) find the largest LMM with a magnitude of ~ 2 $\mu_B$.[9,10] X-ray emission spectroscopy (XES)[11] and photoemission spectroscopy (PE)[12] find values ranging from 1.2 $\mu_B$ to 2.2 $\mu_B$ for 1111 (e.g. CeFeAsO) and 122 (e.g. $SrFe_2As_2$) parents, respectively, while conventional magnetic probes such as neutron scattering, Mössbauer, and muon spin rotation (μ-SR)[13] find the lowest values ranging from 0.9 $\mu_B$ for $BaFe_2As_2$ to ~ 0.3 $\mu_B$ for PrFeAsO. These discrepancies can be attributed to the rapid temporal fluctuations of the local Fe moments, which enable only the techniques with the fastest response (PES, XES) to capture the instantaneous local moments, while conventional, slower techniques provide a time-averaged value. The situation is similarly confusing when describing the evolution of the LMM and electronic states with doping. There is no agreement among both theoretical and experimental studies on whether Co substitutions actually dope carriers in 122 arsenides. Although angle resolved photoemission spectroscopy (ARPES) experiments clearly show modifications of the Fermi surface consistent with charge doping,[14] X-ray absorption spectroscopy experiments reveal no change in the electronic occupation of Fe ions, and deduce that superconductivity is solely induced by bonding modifications.[15]

Some of the discrepancies found in the literature can be explained by realizing that FBS display different degrees of inhomogeneities, depending on details of growth methods and conditions. Inhomogeneities can include interstitials or vacancies forming either ordered or disordered clusters, strain, nanoscale variations in the dopant concentrations, and structural distortions, which have all been observed in FBS.

In this Letter, we report the local Fe magnetic moment across FBS families, free of the effects of extrinsic inhomogeneities, obtained in real space, with sub-nanometer resolution, and time-response as fast as electron dynamical processes (~$10^{-15}$ s), using EELS in the aberration-corrected STEM at room temperature. This extensive study involves data collected over four years on more than 50 TEM specimens.

We find that the local, room-temperature Fe magnetic moment varies significantly across the different families, reaching maximum values of ~ 3.3 $\mu_B$ in the AFS and minimum values of ~ 1 $\mu B$ in the 1111 pnictides. Details of the Fe spectra indicate that the observed variations in LMM originate from different electron occupancies of the Fe 3$d$ orbitals in different compounds. Surprisingly, the dependence of the local magnetic moment on doping is not monotonic as one would expect. A detailed analysis of the Ba(Fe$_{1-x}$Co$_x$)$_2$As$_2$ (0 ≤ $x$ ≤ 0.16) system shows that the LMM drops significantly with increasing doping level at low Co concentrations, exhibiting a minimum close to the onset of superconductivity (x ≤ 0.04). However, for larger Co concentrations, the LMM increases showing a dome-like dependence with a maximum similar to $T_C$ vs doping. Finally, we show direct evidence that the Fe 3$d$ band filling increases as Co concentration increases in Ba(Fe$_{1-x}$Co$_x$)$_2$As$_2$, and therefore Co substitution for Fe does introduce carriers at the Fermi level in FBS.

**Figure 1** is a collection of HAADF images from members of the 1111, 122, 11 (e.g. FeSe), and AFS families in different projections. These images show the exact regions from which the spectra were acquired (immediately before and/or after the images were taken) demonstrating the pristine conditions of the samples for which the magnetic moment was measured. The [100] projections (Figure 1a and 1d) clearly resolve the FeAs and FeSe layers common to all families, with the Fe columns appearing dimmer than the heavier As(Se) columns. In TlFe$_{1.6}$Se$_2$, ordered Fe vacancies arranged to yield a $\sqrt{5} \times \sqrt{5}$ superstructure[16] are directly visible at specific projections, such as the [110] projection shown in Figure 1e.

**Figure 2a** shows the Fe-L edge acquired for metallic Fe, and different FBS and parent compounds. The local Fe magnetic moments were calculated by obtaining the ratio between the intensity of the L$_3$ and L$_2$ peaks (known as L$_{2,3}$ ratio) in the acquired Fe-L edge (see SI for details). L$_3$ and L$_2$ peaks arise from excitations of Fe $2p_{3/2}$ and $2p_{1/2}$ core electrons to unoccupied Fe $3d$ states, respectively. The detailed shape, position and intensity ratio of the L$_3$ and L$_2$ lines depends on several many-body effects and the ion's chemical environment. Specifically, the Fe L$_{2,3}$ ratio is sensitive to 1) Fe oxidation state, 2) Fe spin state, 3) crystal field and local bonding.[17,18,19] Given the dependence on multiple parameters, the L$_{2,3}$ ratio is not typically used to estimate local magnetic moments in transition metal compounds. However, we show here that, in agreement with earlier studies of the L$_{2,3}$ ratio and LMM in Fe- and other transmission-metal-compounds by EELS and x-ray absorption spectroscopy,[20-24] in the case of FBS and their parents, Fe-L$_{2,3}$ spectra give a direct estimate of changes in the local Fe magnetic moment owing to the existence of a linear relationship between L$_{2,3}$ ratio and LMM.

**Figure 2b** shows a plot of the L$_{2,3}$ ratio for some of the compounds studied as a function of the local fluctuating Fe magnetic moment as reported in Refs. 11, 12, and 25 for the same compounds. The Fe L$_{2,3}$ ratio was measured for both [001] projections and [100] projections, and the two measurements gave almost identical values with the exception of the more anisotropic 1111 materials, for which the Fe $L_{2,3}$ ratios for the [100] projections were usually smaller than those for the [001] projection. Reported here are the average values of the measurements in both crystallographic directions. The spectra were collected averaging over regions of a few $nm^2$, and were insensitive to sample orientation and beam channeling, as it was experimentally verified (see the SI for more details).

The plot in Figure 2b indicates that the Fe $L_{2,3}$ ratio has a simple linear dependence on the Fe LMM in the high-temperature paramagnetic state. This empirical relationship finds a justification in the similar chemical and structural environment of the Fe ions in all the Fe-based compounds (characterized by analogous FeAs and FeSe(Te) planes with Fe in the same coordination and formal oxidation state), which leaves the spin state as the main variable affecting the Fe $L_{2,3}$ ratio, while crystal field and bonding effects produce negligible variations. This result is significant because it allows us to estimate Fe LMMs in FBS with sub-nanometer spatial resolution using EELS at room temperature (RT). We note from Figure 2b that the Fe $L_{2,3}$ ratio shows the smallest values for 1111 and 122 materials, intermediate values for the 11s and largest values for the AFSs, providing further experimental evidence of variation in LMM across the FBS families.

In addition to being a measure of the LMM, when summed and properly normalized, the integrated intensities of $L_2$ ($I_{L2}$) and $L_3$ ($I_{L3}$) provide an accurate estimate of the total number of holes in the $3d$ band.[26] Information about the distribution of the Fe $3d$ electrons is obtained considering the spin orbit interaction, which causes a small splitting of the final Fe $3d$ states with total angular momentum $j$, separating the total density of final states in partial densities relative to $j = 3/2$ and $j = 5/2$. Within the dipole approximation, one can derive an expression for the ratio of holes in $j = 5/2$ and $j = 3/2$ states given by[21]

$$\frac{h_{5/2}}{h_{3/2}} = \frac{1}{6}\left[\frac{5 I_{L_3} \omega_{L_2}}{2 I_{L_2} \omega_{L_3}} - 1\right],$$

where $\omega_{L3}$ and $\omega_{L2}$ are the respective transition energies. By calculating the area under the Fe $L_3$ and the Fe $L_2$ peaks, it is thus possible to estimate the number of electrons in the Fe $3d$ band and obtain information about their distribution.

**Figure 3a** shows a plot of the ratio of holes in the $j = 5/2$ and $j = 3/2$ levels ($h_{5/2}/h_{3/2}$) and the total number of holes in the Fe $3d$ band as function of the Fe LMM for all the Fe-based families studied. The hole ratio follows the same trend as the Fe $L_{2,3}$ ratio across the families of Fe-based compounds with $(h_{5/2}/h_{3/2})1111 \leq (h_{5/2}/h_{3/2})122 < (h_{5/2}/h_{3/2})11 < (h_{5/2}/h_{3/2})$AFS. However, the total number of holes does not correlate with the Fe LMM, being nearly constant in all the Fe-based compounds and equal to $\approx 4$. Figure 3a suggests that the difference in the Fe local magnetic moment among the different Fe-based compounds derives mainly from a varying distribution of the same number of electrons within the Fe $3d$ orbitals. The data are in agreement with recent theoretical calculations from Yin *et al.*,[27] which show that in going from the AFS to the 1111 in the order illustrated here, the distribution of electrons changes as electron transfer occurs from the $e_g$ to the $t_{2g}$ orbitals, leaving the total Fe $3d$ occupancy unchanged.

As shown in the schematic of **Figure 3b**, the $L_3$ edge is generated by electronic transitions from filled $2p_{3/2}$ to empty $3d_{5/2}$ orbitals, while $L_2$ corresponds to transitions from filled $2p_{1/2}$ to empty $3d_{3/2}$ orbitals (dipole selection rules). Because the spin-orbit splitting is much larger for the $2p$ levels than the $3d$ levels, the $L_3$ always occurs at a lower energy than the $L_2$, independent on the actual distribution of the $3d$ orbitals, dictated by the combination of crystal field, electron correlations, and spin-orbit effects. The behavior of the hole ratio in Figure 3a indicates that the number of holes in the $3d_{5/2}$ orbitals increases with respect to the number of holes in the $3d_{3/2}$ orbitals in going from the 1111s to the AFSs. Therefore, the occupancy of the $3d_{5/2}$ orbitals progressively decreases from the 1111s to the AFSs but the total number of electrons in the $3d$ orbitals remains nearly constant. The explicit expressions of $3d_{3/2}$ and $3d_{5/2}$ orbitals in terms of $t_{2g}$ ($d_{yz}$, $d_{zx}$, $d_{xy}$), and $e_g$ ($d_{3z^2-r^2}$, $d_{x^2-y^2}$) orbitals can be calculated using the Clebsch-Gordan coefficients (as done in the SI) and the result shows that both $3d_{3/2}$ and $3d_{5/2}$ states are composed by 40% $e_g$ states and 60% $t_{2g}$ states. Therefore, the hole ratio we find cannot give direct information on the occupation of the specific orbitals; however, it can be used to test theoretical models of orbital occupation in different iron-based families. A calculation of the energy evolution of the $j=5/2$ and $j=3/2$ states from the limit of weak crystal field to that of strong crystal field for tetrahedral crystal fields (see SI) shows that $j = 5/2$ and $j = 3/2$ evolve separately into $t_{2g}$ and $e_g$; as shown in the Energy scheme of Fig. 3b.

A major concern in determining the relationships between LMM, band filling, and doping in Fe-based compounds is the ability to recognize and avoid the effects of structural and chemical inhomogeneities, which can only be done using spectroscopic techniques having nanoscale spatial resolution. In fact, STEM/EELS measurements show that the local Co concentration in Ba(Fe$_{1-x}$Co$_x$)$_2$As$_2$ crystals can change significantly on the nanometer scale depending on processing conditions. In addition, stacking faults are observed in cross section, which can also influence the measured electronic response (see SI). The simultaneous acquisition of high-resolution images and atomic-resolution EEL spectra in the STEM guarantees that contributions from defects, secondary phases and doping inhomogeneities can be excluded from the data. In order to explore the dependence of the Fe LMM on doping, images, spectra, and spectrum images were collected for a set of Ba(Fe$_{1-x}$Co$_x$)$_2$As$_2$ single crystals with $x$ ranging from 0 to 0.16. The results are reported in **Figure 4**, as a plot of the Fe LMM as a function of the Co concentration derived by EELS using the same spectra from which the Fe L$_{2,3}$ ratio was calculated. The homogeneity of the Co concentration in the region from which the EEL spectra were acquired was determined by subsequently acquired spectrum images.

Figure 4 shows that the Fe LMM in the paramagnetic state rapidly decreases as compared to the moment of the parent compound for small Co additions in the range $0 < x \leq 0.04$. However, for Co concentrations above 4% ($T_C \approx 10$ K), the magnetic moment starts to increase, reaching a maximum near optimal doping, after which it decreases again as the Co concentration increases further. On the other hand, the total number of holes plotted in Figure 4b decreases, indicating a progressive filling of the Fe $3d$ bands with increasing Co concentration. Figure 4b is a clear indication that Co substitution contributes electrons to the Fe states above the Fermi level, which is in agreement with the notion of electron doping.

The LMM in optimally doped Ba(Fe$_{1-x}$Co$_x$)$_2$As$_2$ is about 30% smaller than in the parent compound. This difference is smaller than what found for Sr(Fe$_{1-x}$Co$_x$)$_2$As$_2$ in Ref. 11. The reason could be related to doping inhomogeneity, which could yield a smaller LMM upon averaging, or might indicate that the maximum of the dome for Sr(Fe$_{1-x}$Co$_x$)$_2$As$_2$ is shifted towards higher Co concentrations. These finding are consistent with a recent study by resonant inelastic X-ray scattering, on optimally-doped superconducting Ba$_{0.6}$K$_{0.4}$Fe$_2$As$_2$, in which spin fluctuations as intense as in BaFe$_2$As$_2$ were measured.[28] Further agreement is found in ARPES,[13] resistivity,[28] Hall coefficient,[29,30] and thermoelectric power experiments, which show large Fermi surface modifications at small electron doping, just before the superconducting transition. As the Fermi surface reconstructs and the $3d$ orbitals shift with respect to the Fermi level as a function of doping, it is plausible that the dynamic Fe moment increases for $x > 0.04$ as a result to a transition to a higher spin state. Density of States calculations show[31] that in 122 arsenides the $d_{3z^2-r^2}$ orbital is nearly empty at the Fermi level, while the $d_{x^2-y^2}$ is mostly occupied. The other three $d_{xy}$, $d_{yz}$, $d_{zx}$ show intermediate occupancy. Based on these calculations, we can focus our attention only on the $d_{3z^2-r^2}$, $d_{xy}$, $d_{yz}$, and $d_{zx}$, assuming that $d_{x^2-y^2}$ is full. We consider three of the orbitals (e.g. $d_{xy}$, $d_{yz}$, $d_{zx}$) as nearly degenerate and the other one ($d_{3z^2-r^2}$) to have a slightly higher energy. A possible explanation for the increase in LMM would be that a tendency to pair develops in the upper energy orbital. As a consequence, it may become energetically advantageous for the occupancy of this orbital to increase, giving rise to a higher spin state (see SI for a schematic model). Although there is evidence that superconductivity might not necessarily be driven by Fermi surface nesting, this tentative explanation certainly needs a more robust support from more sophisticated many-body calculations, which are beyond the scope of this paper. This "unoccupied" orbital may also be a band of states physically located somewhere else such as at the As atoms, that develops with doping. In fact, Co doping has shown to induce a change in the Fe-As bond distances,[32] which has been directly correlated to a change in local Fe magnetic moment.[33]

In conclusion, both the Fe local magnetic moment and Fe orbital occupations of FBS materials can be readily unveiled through the real space capability of aberration-corrected STEM. Bulk probes may miss subtle structural variations, making interpretation difficult or impossible. We have found that Fe 3$d$ orbital occupancy is the key factor controlling the local Fe magnetic moment. In addition, although the ordering of Fe moments needs to be suppressed for superconductivity to arise, the local, fluctuating Fe magnetic moment is a key ingredient for the appearance of superconductivity. These are important steps towards an improved microscopic description of the appearance of superconductivity in Fe-based compounds.

**Experimental Section**

Simultaneous spectroscopy and high angle annular dark field (HAADF) imaging was conducted on nearly 50 samples at room temperature. The 11 chalcogenide, 122 arsenides, and AFS (with the exception of $Li_{0.5}Fe_2Se_2$) were single crystals grown out of self-flux.[34,35] 1111 samples are polycrystalline and were synthesized using a solid-state reaction.[36] Polycrystalline $Li_{0.5}Fe_2Se_2$ samples were produced by a liquid ammonia solvothermal technique.[37]

All single crystal samples were cleaved or cut and polished to about 20 μm in thickness using a precision polishing machine, and subsequently thinned to electron transparency (~20 nm) by Ar ion milling at a voltage ≤ 3 kV with a final cleaning step at 0.5 kV. The microscope used for this study was a Nion UltraSTEM 100™ operating at 100 kV[38] equipped with a second generation 5$^{th}$ order probe aberration corrector, a cold field emission electron gun, and a Gatan Enfina EEL spectrometer. This microscope routinely achieves a spatial resolution of 0.8 Å and has a maximum energy resolution of 300 meV. In some cases, a Nion UltraSTEM 200™ microscope operating at 200 kV,[39] was used, which is capable of higher spatial resolution as compared to the 100 kV electron microscope.

**Supporting Information**
Supporting Information is available from the Wiley Online Library or from the author.

**Acknowledgements**
Research was supported by the Materials Sciences and Engineering Division Office of Basic Energy Sciences, U.S. Department of Energy. Part of the work was supported by ORNL's Shared Research Equipment (ShaRE) User Facility Program, which is sponsored by the Office of Basic Energy Sciences, U.S. Department of Energy. J-CI and WZ acknowledge support from NSF grant No. DMR-0938330.

Received: ((will be filled in by the editorial staff))
Revised: ((will be filled in by the editorial staff))
Published online: ((will be filled in by the editorial staff))

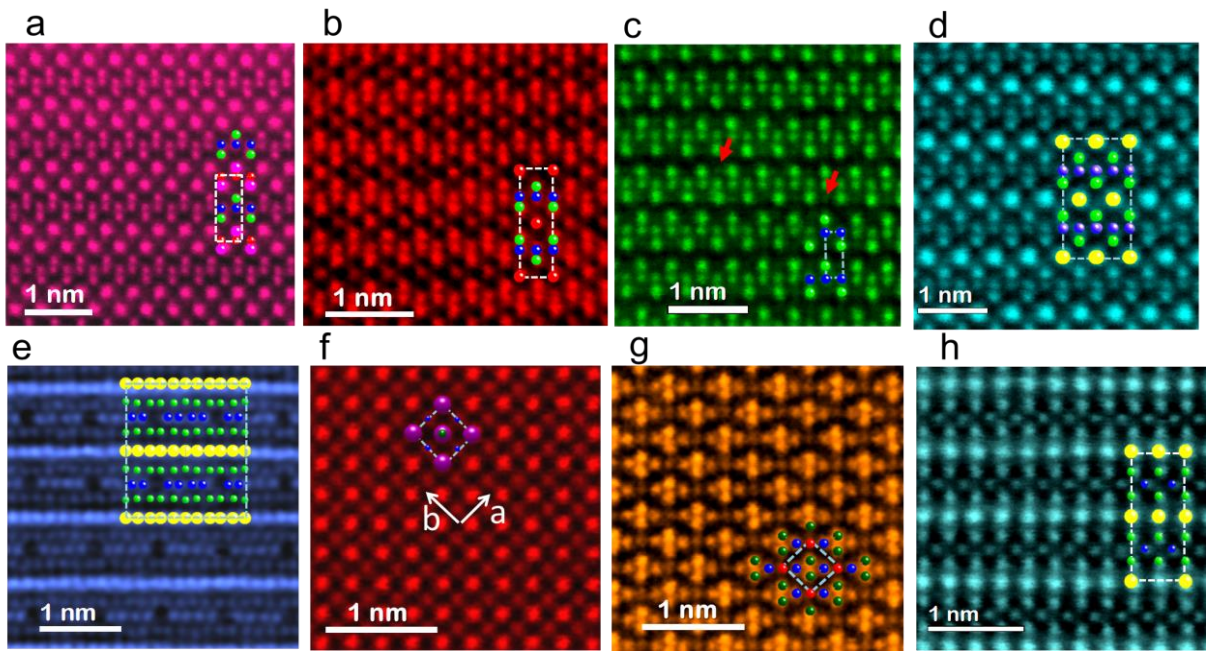

Figure 1. High angle annular dark field (HAADF) STEM images of various FBS in different projections. a) PrFeAsO viewed along the [100] projection. Pr atoms are drawn in pink and O atoms are drawn in red. For all the drawings blue spheres indicate Fe atoms and green spheres indicate As atoms. b) $Ca_{0.85}Pr_{0.15}Fe_2As_2$ ($T_C$ = 45 K) [100] projection. Ca atoms are red. c) $Fe_{0.99}Te_{0.45}Se_{0.55}$ ($T_C$ = 14 K ) [100] projection. Te and Se atoms are green. Red arrows indicate interstitial Fe atoms. d) $TlFe_{1.6}Se_2$ having fully ordered Fe vacancies, [100] projection of the $ThCr_2Si_2$ subcell. Tl is yellow, Se green. e) $TlFe_{1.6}Se_2$ having fully ordered Fe vacancies, [110] zone-axis of the vacancy-ordered $\sqrt{5} \times \sqrt{5}$ supercell ([310] of the $ThCr_2Si_2$ subcell). Vacant Fe columns appear as dark spots. f) $BaFe_{1.92}Co_{0.08}As_2$ ($T_C$ = 24 K), [001] projection. Ba is purple. g) $CaFe_2As_2$ [111] projection. h) $TlFe_{1.6}Se_2$ with disordered Fe [110] projection of the $ThCr_2Si_2$ cell.

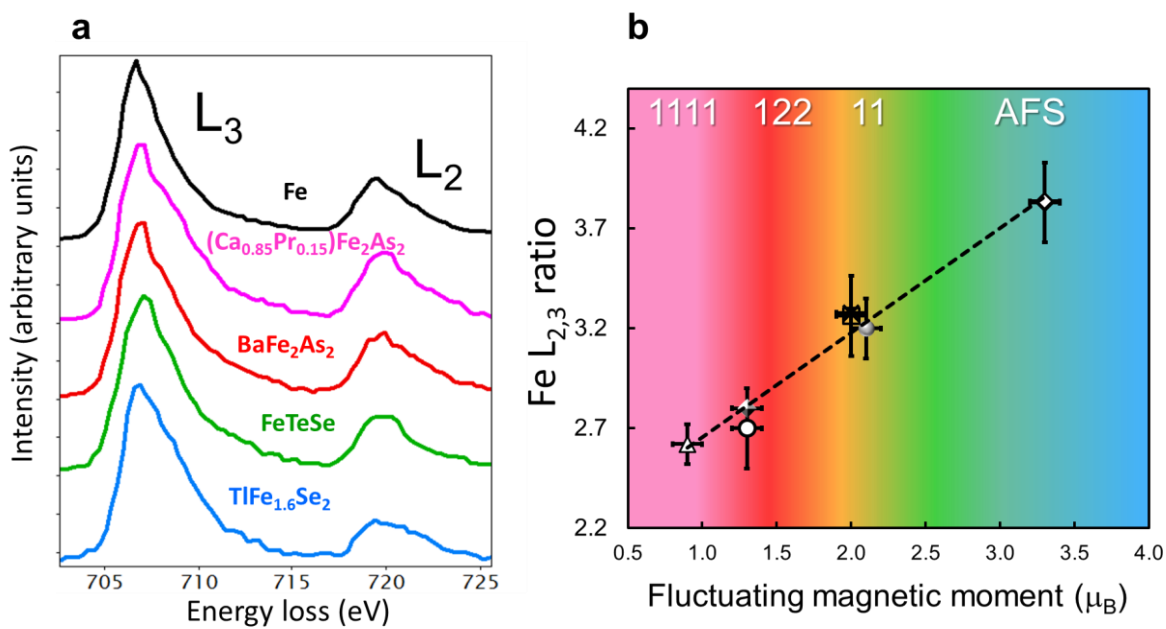

Figure 2. The Fe-$L_{2,3}$ ratio and its dependence on the local fluctuating Fe magnetic moment in FBS. a) comparison of Fe $L_{2,3}$ edges for Fe metal (black curve) and some of the compounds studied and plotted in b). b) plot of Fe $L_{2,3}$ ratio vs. Fe LMM measured in Refs. 10, 11, 25. The plotted moment for FeSe and $Fe_{0.99}Te_{0.45}Se_{0.55}$ were taken both equal to 2 $\mu_B$ (the value reported in Ref. 10 for $Fe_{1.2}Te$, $Fe_{1.03}Te$, and $FeTe_{0.3}Se_{0.7}$) because the whole series $FeTe_xSe_{1-x}$ $0 \leq x \leq 1$ is found to have similar LMMs (see Ref. 25). Data points were obtained by multiple measurements. From left to right: $(Ca_{0.85}Pr_{0.15})Fe_2As_2$ (△), CeFeAsO (○), $BaFe_2As_2$ (◆), FeSe (×), $Fe_{0.99}Te_{0.45}Se_{0.55}$ (□), $SrFe_2As_2$ (●), and $TlFe_{1.6}Se_2$ (◇).

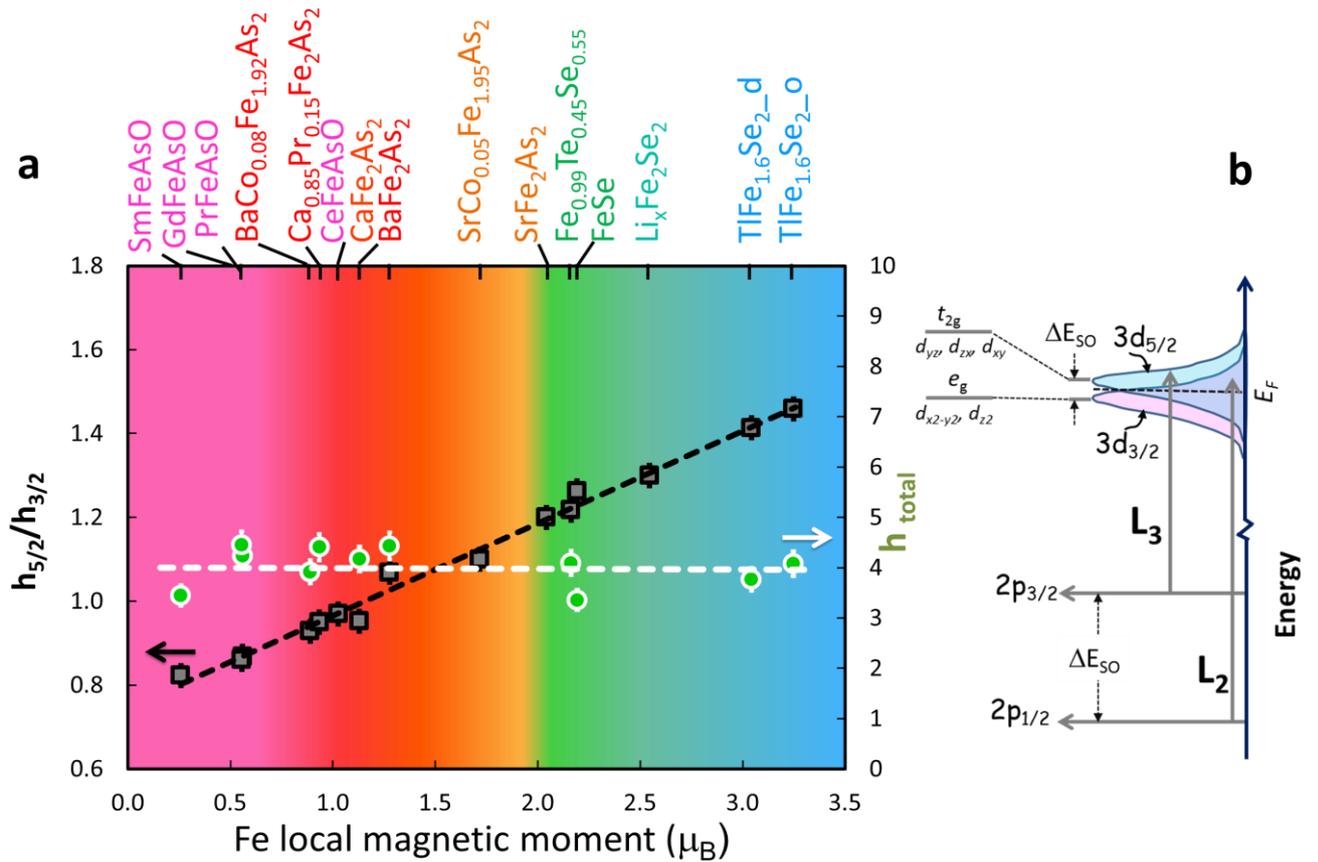

Figure 3. The ratio of holes in $j = 3/2$ and $j = 5/2$ (squares) and the total number of holes in the 3$d$ band (circles) vs. $L_{2,3}$ ratio. a) The ratio of holes increases according to the sequence $(h_{5/2}/h_{3/2})1111 \leq (h_{5/2}/h_{3/2})122 < (h_{5/2}/h_{3/2})11 < (h_{5/2}/h_{3/2})$AFS, indicating that different FBS families have different distributions of electrons in the Fe 3$d$ orbitals. The total number of holes fluctuates around a value of $\approx 4$. Subscripts "o" and "d" for TlFe$_{1.6}$Se$_2$ refer to the compounds with ordered and disordered vacancies, respectively. b) schematic of electronic transitions and atomic orbitals probed by the Fe-$L_{3,2}$.

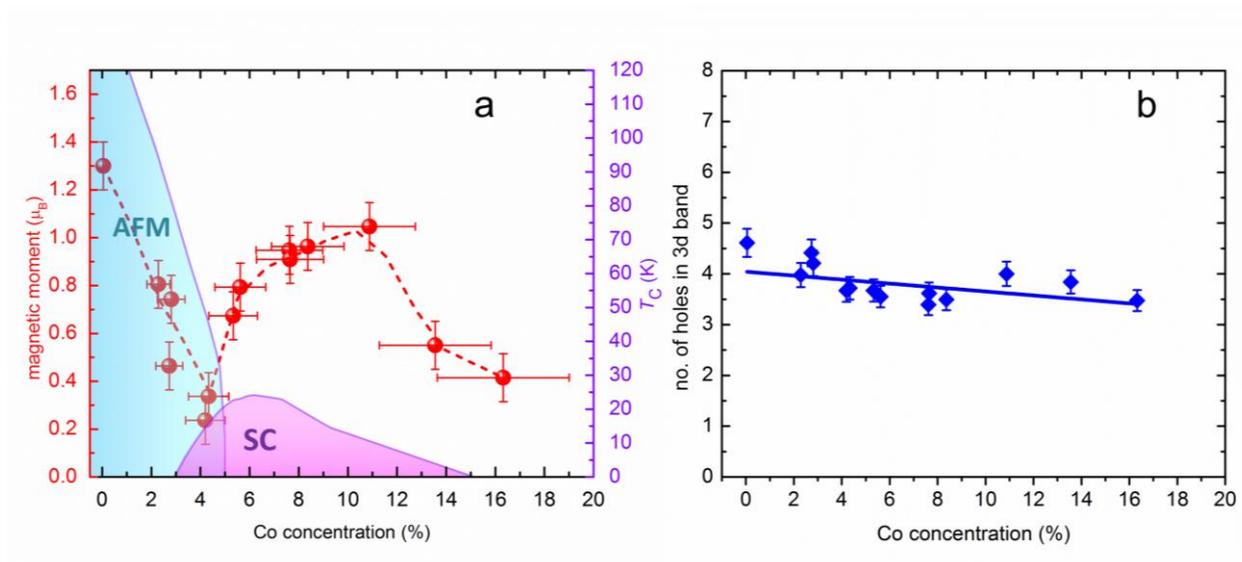

Figure 4. Local Fe magnetic moment and total number of holes in the Fe 3$d$ orbitals in Ba(Fe$_{1-x}$Co$_x$)$_2$As$_2$ as function of local Co concentration. a) the EELS-derived Fe magnetic moment and overlaid Ba(Fe$_{1-x}$Co$_x$)$_2$As$_2$ phase diagram adapted from Ref. 40. b) the number of holes in the Fe 3$d$ orbitals decreases as the Co concentration increases.

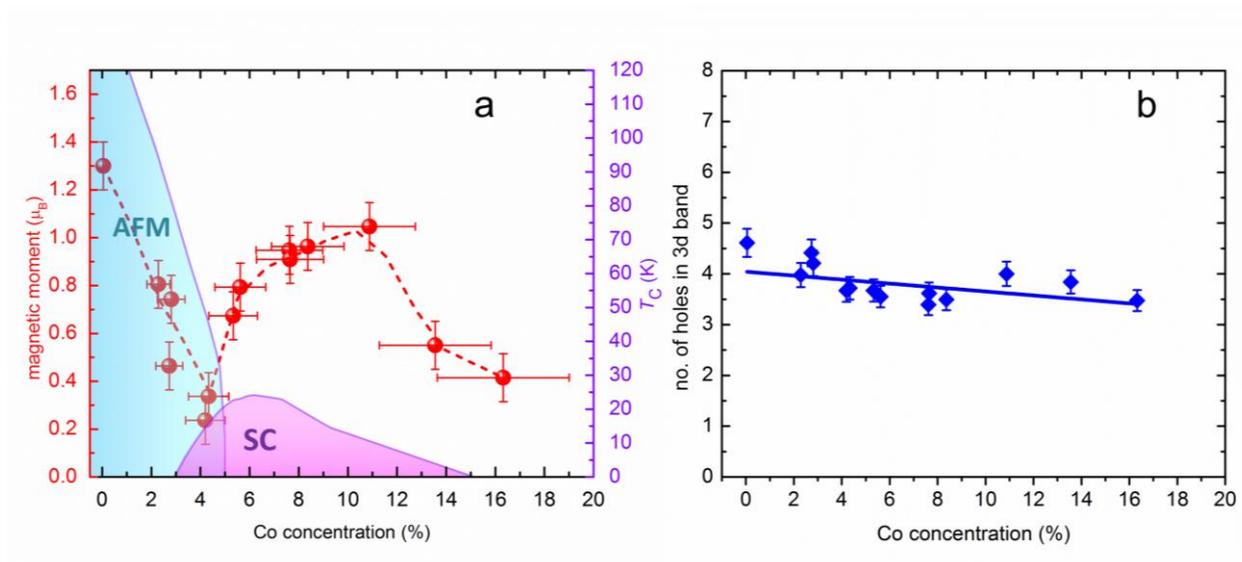

Figure 4. Local Fe magnetic moment and total number of holes in the Fe 3$d$ orbitals in Ba(Fe$_{1-x}$Co$_x$)$_2$As$_2$ as function of local Co concentration. a) the EELS-derived Fe magnetic moment and overlaid Ba(Fe$_{1-x}$Co$_x$)$_2$As$_2$ phase diagram adapted from Ref. 40. b) the number of holes in the Fe 3$d$ orbitals decreases as the Co concentration increases.

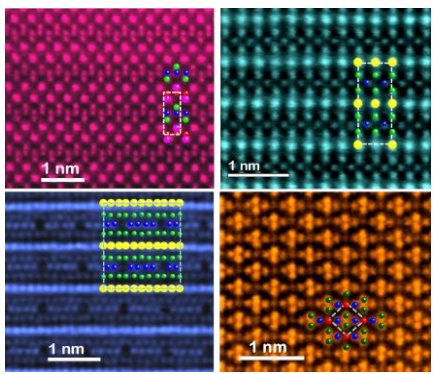

ToC figure
The intrinsic Fe local magnetic moment and Fe orbital occupations of iron-based superconductors are unveiled through the local, real-space capability of aberration-corrected STEM/EELS. Although the ordering of Fe moments needs to be suppressed for superconductivity to arise, the local, fluctuating Fe magnetic moment is enhanced near optimal superconductivity.

Supporting Information

**Orbital occupancy and charge doping in iron-based superconductors**

*Claudia Cantoni\*, Jonathan E. Mitchell, Andrew F. May, Michael A. McGuire, Juan-Carlos Idrobo, Tom Berlijn, Matthew F. Chisholm, Stephen J. Pennycook, Athena S. Sefat, and Brian C. Sales*

**This file includes:**

Detailed Methods

Additional Figures S1 to S6

Additional Reference

Samples were prepared in plane-view geometry (for viewing along <001>) and in cross-section geometry (for viewing along equivalent <100> and <010> directions). For the latter geometry, thin crystal platelets were embedded in high temperature epoxy (cured at 120° C) for support during polishing, given the soft and malleable nature of the crystals. As prepared TEM samples were heated up to 80 °C in high vacuum ($10^{-7}$ Torr) for several hours in order to remove hydrocarbon contamination from air exposure, and then rapidly loaded in the microscope.

The samples were imaged using the high angle annular dark field detector, which selects diffracted electrons that have undergone elastic scattering in close proximity to the nuclei (and are therefore scattered at high angles from the optic axis) yielding an intensity nearly proportional to $Z^2$ ensuring chemical information (heavier columns appear brighter). The probe forming aperture semiangle was 30 mrad.

For the EEL spectra, a dispersion of 0.3 eV/channel was used together with an acquisition time of 5-10 s. Typically, ten or more spectra were acquired and averaged together. In addition, spectrum images were acquired with a dwell time of 0.5-1 s and the spectra within the image were subsequently averaged. EELS is a fast process, where the electrons take only a few femto seconds ($10^{-15}$ s) to travel through and interact with the specimen. Thus, every single electron collected in the EELS experiments provides a snapshot of the magnetic moment. However, since a large quantity of electrons need to be collected to increase the signal to noise ratio of the spectra, long acquisition times (few seconds) compared to the electronic excitation processes are required. For all the oxygen-free samples studied, oxygen contamination was not observed, as revealed by the absence of an O-K edge at 530 eV.

The ELL spectra were processed in the following way: The background was removed using a power law function (see Fig. S1), plural scattering effects were removed by deconvolution of the low loss spectrum using the Fourier-ratio method [1] implemented in Digital Micrograph, and the contribution from transitions to the continuum was removed using a Hartree-Slater-type step function, as illustrated in Fig. S2. Then, the Fe $L_{2,3}$ ratio was obtained by dividing the $L_3$ integrated intensity by that of the $L_2$, both calculated using an energy window of 10 eV starting at the edge onset (Fig. S2). The number of holes in the Fe $3d$ band was obtained by adding the two areas and then dividing by the area in a normalization window 33 eV in width beginning 33 eV past the $L_3$ white-line onset. The value obtained was then compared to the same value obtained for Fe metal with known number of holes in the 3d bands. For the majority of samples, the EEL spectra were collected from sample regions having a thickness in the range 0.3 – 0.5 $\lambda$ ($\lambda$ = electron mean free path), as calculated using the log-ratio method [S1]. However, in certain instances it was not possible to find regions of similar thickness and the data were acquired from regions with much smaller or larger thickness. Although the spectrum was deconvoluted, the change in thickness largely affected the region of the spectrum past the $L_3$ used for normalization when calculating the total number of holes; therefore these data were not considered in the analysis of $h_{tot}$.

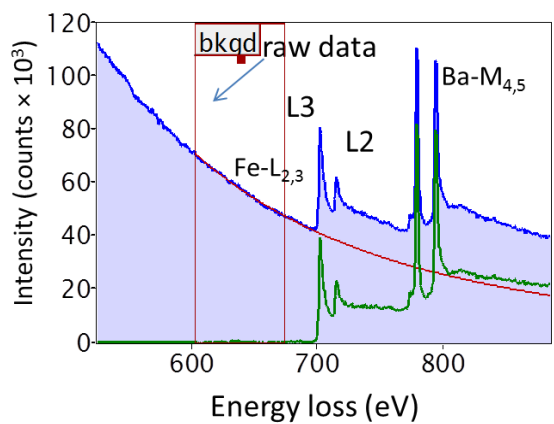

**Fig. S1**. EELS raw data

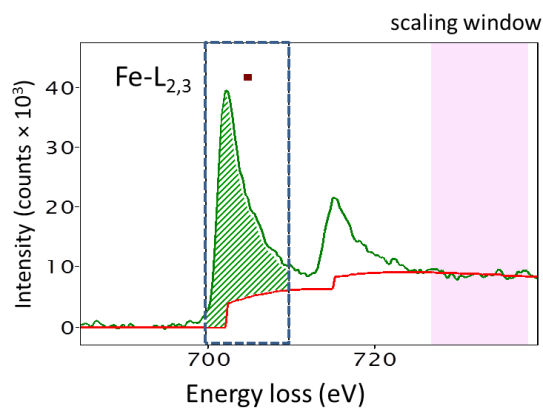

**Fig. S2** step function and windows used in the data analysis

We have collected spectra for several orientations of the sample away from strong diffracting conditions and compared them to the on-axis spectra, all other experimental conditions being equal. We found that there is no difference between the spectra. Moreover, the calculated Fe $L_{2,3}$ ratios showed a standard deviation of only 0.05, well within the errors reported in the manuscript. Figure S3 shows the Fe L-edge from $BaFe_2As_2$ acquired with the beam along the [001] and in an off-axis condition, respectively, as illustrated by the corresponding images and Ronchigrams. The two Fe-L edges are very similar and overlap when normalized to their maximum intensity. At the same time, it is clear that the Fe-$L_{2,3}$ spectra for $BaFe_2As_2$ (on- and off-axis) differ significantly from other iron-based compounds shown in the same graph for comparison. The EELS collection semiangle was 48 mrad.

We did not rely on the SEM/EDX measurements for the local Co composition because in some samples the Co concentration had been observed to vary significantly with position (Fig. S3). The Co concentration was obtained by creating a calibration plot of the ratio between the Co-$L_3$ and the Ba-$M_4$ peak intensities from the EEL spectra versus the EDX Co concentration for a few samples showing uniform Co distribution, as shown in Fig. S4. The data points were fitted using a polynomial function and the Co concentration for all other samples was then calculated by interpolation/extrapolation.

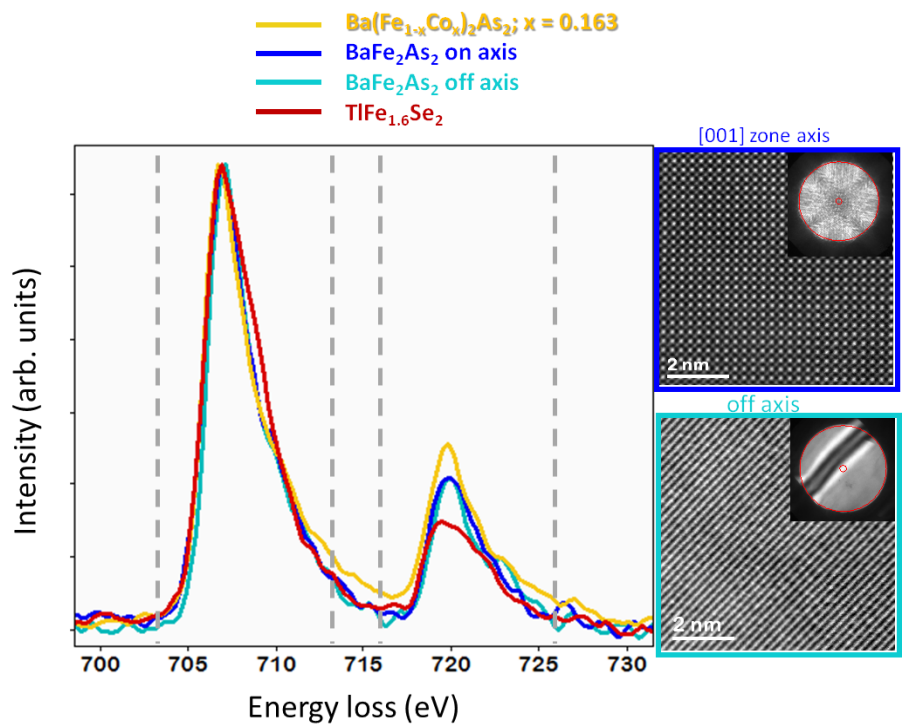

**Fig. S3**. Fe-$L_{2,3}$ edges from $BaFe_2A_2$ (acquired with the beam along the [001] and in an off-axis condition), $TlFe_{1.6}Se_2$, and $Ba(Fe_{1-x}Co_x)_2As_2$; x = 0.163.

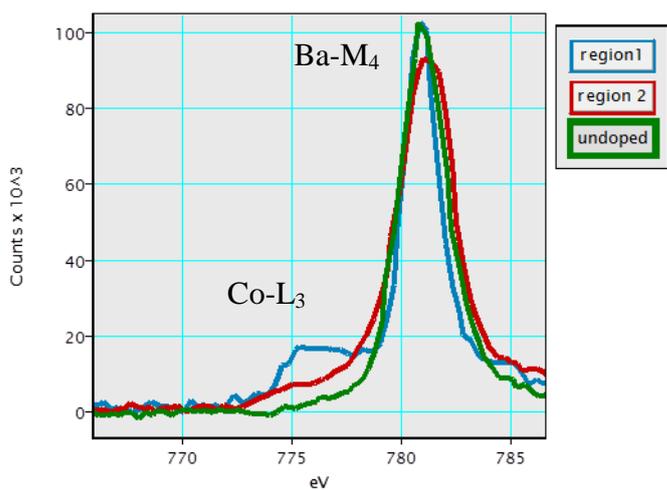

**Fig.S4.** Co/Ba ratio in different region of same sample compared to undoped crystal

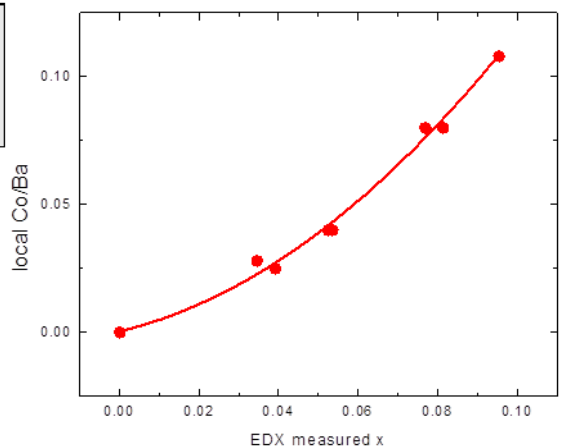

**Fig.S5.** Co concentration calibration curve

As mentioned in the main text, staking fault-type of defects as the ones shown in Fig. S5 were occasionally observed. EEL spectra were in all cases acquired well away from these defects, in pristine region of the sample

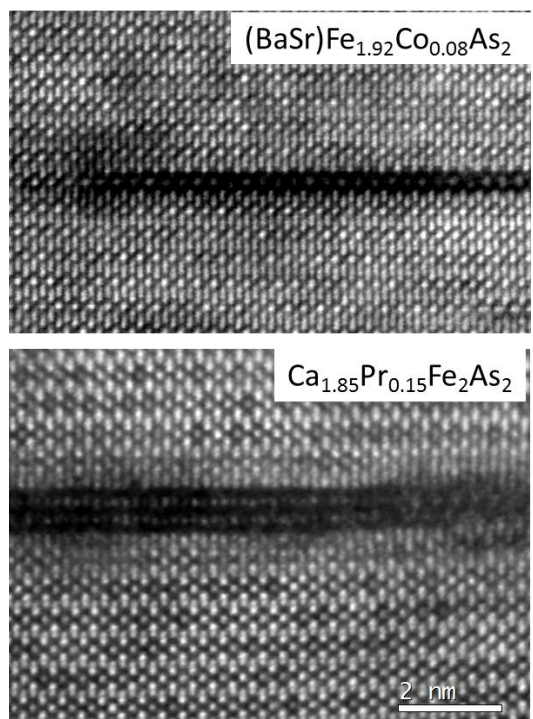

**Fig. S6** Example of defects occurring in 122 arsenides

The procedure we used to select the regions is illustrated in the figure below (Fig. 1).

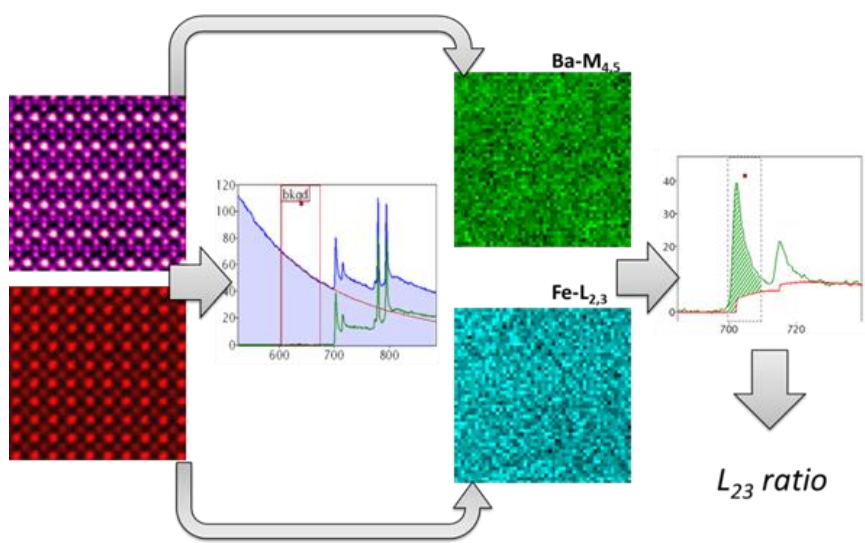

**Fig. S6**. Schematic of the procedure used to insure data in Fig. 4 were acquired for homogeneous regions.

We first selected several structurally flawless regions and acquire high-quality spectra (with high signal-to-noise ratio) by averaging on the entire region, then we went back and performed a spectrum image on the same region to find out whether the Co (or Fe) concentration was uniform. We selected the measurements done on homogeneous regions.

Schematic of electron transition resulting in an increase of LMM
Let's consider two nearest neighbors Fe atoms both in S=1 configuration, such as:

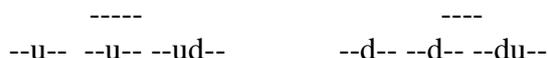
    -----                  ----
--u--  --u--  --ud--        --d--  --d--  --du--

Let's assume that there is a pairing tendency at the empty orbital that favors the formation of singlets. Then we can take the "d" of the left and the "u" of the right, form a singlet in the upper part as follows

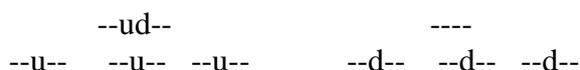
      --ud--               ----
--u--  --u--  --u--        --d--  --d--  --d--

Now we have two S=3/2 states.

The "pair" above can be in one site or the other. Then pairing can lead to the appearance of S=3/2 states from the S=1 original states. Of course, when we refer to empty states, we do not mean that these need to be totally empty, just partially.

[S1] R. F. Egerton Electron energy loss spectroscopy in the transmission electron microscope. Second Edition, Plenum Press, New York (1996).